# Energy-resolved detection of single infrared photons with $\lambda$ = 8 μm using a superconducting microbolometer

Boris S. Karasik[1,a], Sergey V. Pereverzev[1,b], Alexander Soibel[1],

Daniel F. Santavicca[2], Daniel E. Prober[2], David Olaya[3,c], and Michael E. Gershenson[3]

[1] *Jet Propulsion Laboratory, California Institute of Technology, Pasadena, CA 91109*

[2] *Yale University, New Haven, CT 06520*

[3] *Rutgers University, Piscataway, NJ 08854*

We report on the detection of single photons with $\lambda$ = 8 μm using a superconducting hot-electron microbolometer. The sensing element is a titanium transition-edge sensor with a volume ~ 0.1 μm$^3$ fabricated on a silicon substrate. Poisson photon counting statistics including simultaneous detection of 3 photons was observed. The width of the photon-number peaks was 0.11 eV, 70% of the photon energy, at 50-100 mK. This achieved energy resolution is one of the best figures reported so far for superconducting devices. Such devices can be suitable for single-photon calorimetric spectroscopy throughout the mid-infrared and even the far-infrared.

---

[a] Electronic mail: boris.s.karasik@jpl.nasa.gov

[b] Presently with the Lawrence Livermore National Laboratory, Livermore, CA 94551

[c] Presently with the National Institute of Standards and Technology, Boulder, CO 80305

Single-photon calorimetry using bolometers is becoming a practical spectroscopic technique in X-ray [1,2] and even optical [3,4,5] applications. The small heat capacity is key to achieving the best possible energy resolution, so low-temperature operation of the bolometer is required. Even though semiconductor solid state photomultiplier [6] and quantum-dot [7] devices, and superconducting nanowire detectors [8] have achieved single-photon sensitivity in the IR spectral range, the superconducting transition-edge sensor (TES) is one of a very few devices exhibiting a monotonic (often linear) response to photon energy. It is thus capable of spectroscopy of each photon in a weak photon flux directly, without any external wavelength-dispersive elements. This also enables a photon-number-resolving (PNR) detector for photons of specific wavelength. The ability to provide simultaneously spectral, temporal, and spatial data (if an array of calorimeters is used) would also be useful at longer mid-infrared (mid-IR) and even far-infrared (far-IR) wavelengths. For these spectral ranges, the microbolometer shows promise, as our present work demonstrates. A conventional micromachined (membrane based) calorimeter uses a TES thermometer and separate absorber and thermal link [1,2]. It would need to operate at impractically low cryogenic temperatures in order to have small specific heat, to reach an energy resolution useful in the mid-infrared range. It would also have a very slow response.

In the visible and the near-IR regions, detection of individual photons uses the hot-electron version of TES exclusively [4,5,9,10]. It is denoted a hot-electron bolometer (HEB) even when used as a single-photon calorimeter. Here the superconducting TES is the absorber, thermometer and thermal link, all in one. This minimizes the total heat capacity, which sets the minimum detectable energy. Also, the relaxation of the electron temperature occurs due to the cooling of the hot electrons by emission of phonons. At low temperature this process happens faster than the thermal relaxation in micromachined bolometers via heat conduction through the $Si_xN_y$ membrane. The resulting higher photon count rate of the HEB gives a larger dynamic range for the microbolometer in the photon-counting mode.

In recent years, visible/near-IR ($\lambda \leq 1.5$ μm, photon energy $E_{photon} \geq 0.8$ eV) single-photon HEB detectors have undergone significant development due to their importance for quantum communication applications where PNR, low dark count rate, and high quantum efficiency are required [5,11]. However, these device designs use absorber areas $> \lambda^2$. This approach cannot be used at longer wavelengths, due to the increase of the minimum detectable energy with the larger area. Submicron-size hot-electron bolometers (nano-HEBs) that employ low-loss antennas for coupling in the radiation are predicted to be sensitive to low-energy single photons [12,13] down to THz ($E_{photon} \sim$ few meV) and, possibly, below. In recent work [14], we obtained a minimum resolved energy $\Delta E_{FWHM} \approx 0.1$ eV for a 0.1 μm$^3$ titanium (Ti) microbolometer operating at 300 mK. We used a technique that simulates absorption of a mid-IR photon by a microwave pulse of equivalent energy. In the present work, we study single-photon detection of weak laser pulses with $\lambda = 8$ μm ($E_{photon} = 0.16$ eV) using a Ti microbolometer with a similarly small volume and zero-field critical temperature $T_C = 360$ mK.

The HEB device with the dimensions of 6 μm × 0.4 μm × 56 nm was fabricated on high-resistivity Si (with natural oxide) by means of *in-situ* double-angle e-beam evaporation of Ti and niobium (Nb) in vacuum [13]. The role of Nb is to form contact leads slightly overlapping with Ti (an overlap area $\sim 0.02$ μm$^2$), serving as Andreev barriers to prevent the fast outdiffusion of the electron thermal energy. The much slower electron-phonon relaxation process is the main cooling mechanism in the 50-300 mK temperature range studied. In our previous studies [15,16] we found that the near-equilibrium electron-phonon relaxation time $\tau_{e-ph}$ ranges from 5 μs at 300 mK to 1 ms at 50 mK in similar devices. The initial electron heating after photon absorption is rapidly shared with the other electrons and with high-energy phonons on a fast time scale, < 1 ns [16]. Thus, the electrons in the Ti volume quickly come to an increased temperature $T_e$. The electron temperature profile quickly equilibrates over the device volume since the diffusion time ~ $L^2/D$ = 150 ns ($L$ is the device length, $D$ = 2.4 cm$^2$/s [17] is the Ti electron diffusivity) is much shorter than $\tau_{e\text{-}ph}$. This causes the resistance change on the superconducting transition.

The device was mounted on the cold finger of a dilution refrigerator in a light-tight copper (Cu) box. An in-house developed monochromatic quantum cascade laser (QCL) [18] was used as the source of faint pulses. The laser was mounted in vacuum on the 4-K flange of the dilution refrigerator and the radiation was guided to the HEB device using stainless steel and Cu light-pipes of a few mm diameter. A large attenuation of the pulse energy was achieved by inserting in the light-pipe several small apertures narrowing the opening for the beam to about 1 mm. Stacks of two-three 1-mm thick Teflon and 0.2-mm thick Zitex® G108 [19] sheets were used at the mixing chamber temperature and at the 1-K pot in order to additionally attenuate the 8-µm laser radiation and also to block short-wavelength thermal radiation emitted by the laser or leaking from warm top parts of the dilution refrigerator. Also, Cu foil mesh filters (~30-µm square holes) were placed at different temperatures to block longer wavelength thermal (microwave) radiation. The illumination of the device occurred from the open end of the light-pipe a few mm from the device surface. The device was covered by a Cu shield with a pinhole through which the photons arrived (this helped to reduce the direct heating of the Si substrate by scattered radiation). Thus, only a tiny fraction of the 8-µm photons in the emitted pulse reached the bolometer. After adjustment of the overall attenuation, continuous control of the average number of photons absorbed by the HEB, $\mu$, was obtained from $\mu$ = 0 to $\mu$ > 10 using only adjustments of the QCL pulse duration and amplitude. As many other current-pumped semiconductor lasers, the QCL has a sharp generation threshold where the spontaneous emission turns into a strong coherent emission just within a few percent of the driving current increase [18,20]. Whereas Joule heating of the laser is nearly the same above and below threshold, only thermal emission might be expected below threshold where, indeed, no photons were detected. This comparison allowed us to clearly establish the presence of the radiation with $\lambda = 8$ μm, the only line present in the laser spectrum.

To perform measurements at different temperatures, a small superconducting solenoid with iron core was mounted just outside the Cu box containing the HEB device. The magnetic field created by the solenoid is perpendicular to the Ti film surface and was sufficient to suppress the critical temperature of Ti down to $T_C = 50$ mK. This field was too weak to reduce the blocking efficiency of the Nb Andreev contacts. We found that the temperature dependence of $\tau_{e\text{-}ph}$ was consistent with the electron-phonon relaxation mechanism even at the lowest values of $T_C$, [16] which indicates that the Nb contacts indeed prevented the diffusion cooling and were not perturbed by the applied magnetic field.

The HEB was voltage biased by sending a constant current through a chip resistor ($R_L = 0.33$ Ω) mounted on the mixing chamber. A dc Superconducting Quantum Interference Device (SQUID) mounted on the 1-K pot was connected in series with the HEB device to measure the current, $I$, through the device (see inset in Fig. 1). In order to suppress any unwanted heating effects due to spurious electromagnetic radiation, the bias leads were heavily filtered by RC-filters at both 4 K and the mixing chamber. The SQUID leads were also filtered by a low loss low-pass LCR-filter molded into Eccosorb® CRS-124 silicone absorber to prevent the emission of the noise and/or Josephson radiation from the SQUID into the HEB device. The overall bandwidth of the readout was ~ 70 kHz which was sufficient for passing undistorted photon-induced pulses at 50-150 mK, where $\tau_{e\text{-}ph} > 100$ μs.

Two current-voltage characteristics (IVC) of the device are shown in Fig. 1. The curve taken at 100 mK required a magnetic field to be applied in order to avoid instability which would otherwise occur when the current exceeds the critical value $I_C$. For the 100 mK curve, single-photon detection, as we describe later, occurs only in a narrow voltage range above $V_b = 30$ nV. Both the pulse amplitude and its decay time varied when the bias point was moved from low voltage to large bias voltage, which corresponds to the nearly normal state of a bolometer. The peaked shape of the IVC is typical for TES devices. The maximum responsivity to radiation power occurs when the bias voltage is slightly above the point of maximum current, i.e. where $dI/dV_b$ is negative. Since the energy of a single 8-μm photon can cause significant effect due to the small electron heat capacity of the HEB, we observed some response even when the bias point was below $I_C$. The detection mechanism here could be a mixture of the kinetic inductance response at the earliest and latest stages of the equilibration process and the bolometric resistive response during the middle stage. We did not observe single photon detection under these conditions, though the amplitude of the response was large.

Detection of single or few-photon events occurred in the bias range beginning roughly from the point of maximum response up to the normal state (see Fig. 1). Since the duration of the QCL pulse $\tau_{QCL}$ was much shorter than $\tau_{e\text{-}ph}$, we could trigger the data recording at the start of each emitted photon pulse. This allowed us to obtain photon-counting statistics. The QCL was triggered at a rate of only 20 to 50 Hz due to limits of overall heating of the mixing chamber by the laser radiation. We recorded $10^4$ pulse events at each bias point using a fast oscilloscope card. A low-pass filter with a sharp cut-off at ~ 10 kHz was used for rejecting white noise outside of the HEB bandwidth (Fig. 2).

The processing of the digitized waveforms included averaging all the waveforms to find the averaged pulse waveform and its peak time position. Then the amplitudes of individual traces were determined at this time, and the histogram of these amplitudes was plotted over ~ 100 equal size bins. This provided for a sufficiently smooth yet well resolved histogram of the amplitude count statistics (see Fig. 3). The data of Fig. 3 were taken at 50 mK with the QCL pulse width $\tau_{QCL}$ ranging between 2 μs and 20 μs. Since $\tau_{QCL} \ll \tau_{e\text{-}ph}$ at this low temperature, $\mu$ was proportional to $\tau_{QCL}$. As $\mu$ increased, more photon number peaks become visible. No new photon number peaks are observed after the pulse amplitude reaches 13 nA. This is where the HEB device is fully normal at this temperature and field. The associated current clipping and the pile up of the pulse amplitudes manifest themselves in a narrow peak at 13 nA. The histograms confirm that we detect simultaneously, with some amplitude noise, $k$ = 0, 1, 2, or, possibly, 3 photons per pulse.

The count histogram taken at $T$ = 100 mK is shown in Fig. 4. We use it for the quantitative analysis of the data. The relative height of the histogram peaks follows the Poisson distribution: $h_k = \mu^k \exp(-\mu)/k!$. This also agrees with the data of Fig. 3 for small values of $\mu$. For example, for $\tau_{QCL}$ = 2 μs, the ratio $h_1/h_0 \approx 0.13$, that is $\mu$ = 0.13. Then for $\tau_{QCL}$ = 10 μs, $\mu$ should be ≈ 5×0.13 = 0.65. Indeed, in this case, $h_1/h_0$ = 0.62, and $h_2/h_0$ = 0.25, that is, close to the expected values of $\mu$ and $\mu^2/2$.

In order to describe the broadening of the peaks by noise, we use the following combined Gaussian-Poisson distribution:

$$H(I) = \sum_{k=0}^{3} \exp\left[-\left(I-I_k\right)^2 / 2\sigma^2\right] \mu^k / k! , \qquad (1)$$

where $I_k$ is the average amplitude of counts corresponding to the absorption of $k$ photons. Equation 1 agrees with the data of Fig. 4 when $\sigma$ = 1.2 nA and $\mu$ = 0.47 are used as fitting parameters. The dark count noise is also well approximated by the Gaussian curve with the same value of $\sigma = 1.2$ nA.

To understand the broadening seen in the data of Fig. 4, we consider the current noise spectrum. From Fig. 2, we compute a total rms current noise with $\sigma_I = 1.04$ nA, in good agreement with $\sigma_I = 1.2$ nA derived from the Gaussian fitting of data in Fig. 4 for the dark count noise. We observed single-photon detection with practically unchanged broadening of each photon number peak, between 50 mK and 150 mK. Above 150 mK the contrast between the separated peaks of different $k$ values degraded. The photon number became undistinguishable above 200 mK. We tried to improve the resolution by processing data further with a digital Wiener filter based on the waveform of the averaged response and the noise spectrum of Fig. 2. We did not obtain any significant reduction of the peak widths with this approach.

We now consider the energy resolution of the Ti HEB in this experiment. This parameter is derived from the width of the photon number peaks. The peaks in Fig. 4 are nearly equally separated which means that the output of the bolometer is linear in this energy range. This gives an energy scale of $E_{photon} = 0.16$ eV between adjacent peaks. Thus, the corresponding minimum resolved energy was $\Delta E_{FWHM} = 0.11$ eV. Theoretically, one would expect $\Delta E_{FWHM} = \kappa \left( k_B T_e^2 C_e \right)^{1/2}$, where the prefactor $\kappa$ depends on the effective noise bandwidth. This bandwidth can be much larger than the signal bandwidth [21,22]. A significant reduction of $\Delta E_{FWHM}$ due to negative electro-thermal feedback (ETF) [21] was predicted for TES sensors with a sharp superconducting transition (parameter $\alpha = (T/R)dR/dT \gg 1$). Since we use a magnetic field, the transition broadens and $\alpha$ is not large. In this case, a conservative estimate (the signal bandwidth equals the noise bandwidth) can be made using a general expression for $\Delta E_{FWHM}$ when the noise due to electron thermal energy fluctuations (TEF) dominates [23]: $\Delta E_{FWHM} = \sqrt{4\ln 2}\sqrt{4k_B T_e^2 C_e}$. Since $T_e = T_C \approx 140$ mK and $C_e = \gamma V T_e$ ($\gamma = 310$ J K$^{-2}$ m$^{-3}$ is the normal state Sommerfeld constant, $V$ here is the Ti volume), we predict $\Delta E_{FWHM} = 38$ meV $= 0.24 E_{photon}$.

We find that the experimental $\Delta E_{FWHM}$ is about a factor of three times the value predicted by the bolometric model. We think that the main cause is the inefficiency of the conversion of photon energy into the thermal energy of hot electrons $k_B T_e$ and, consequently, into the current response. A commonly cited reason for that is the loss of the photon energy due to the emission of hot phonons leaving the device volume without being re-absorbed by electrons. For example, in tungsten optical TES on Si, the photon energy efficiency has been found to be $\varepsilon \approx 40\%$ [4,24], whereas this efficiency reaches $\varepsilon \approx 80\%$ when Si$_3$N$_4$ membrane is used as a substrate [25].

In the strong ETF limit, the energy is removed from the TES by a sharp drop of the current on the time scale much shorter than the intrinsic thermal relaxation time ($\tau_{e-ph}$, in our case). Then the corresponding change of the Joule power gives an absolute measure of the energy absorbed in the TES [2,24,25]. Since our HEB device did not operate in the strong ETF limit, we applied a different technique to estimate the photon energy efficiency.

We obtained a series of IVCs at different bath temperatures and recovered $R(T_e)$ from each of them using a heat balance equation:

$$\Sigma V \left( T_e^{n+2} - T^{n+2} \right) = I^2 R(T_e), \qquad (2)$$

where $\Sigma = 6.2\times 10^9$ W m$^{-3}$ K$^{-5.5}$ and $n = 3.5$ are the parameters of the electron-phonon coupling derived from the previously measured magnitude and temperature dependence of $\tau_{e-ph}(T) \sim T^n$ in a similar device [16]. The resulting $R(T_e)$ curves almost coincide (inset in Fig. 5) indicating that $R$ depends only on $T_e$. This confirms the validity of the bolometric model. Note that the transition width is large ($\Delta T_C \approx 30$ mK).

From an $I(T_e)$ plot obtained from the same IVCs (Fig. 5), one can conclude that the dependence of the current on electron temperature is almost linear within the superconducting transition range. This explains the equidistant positions of the photon numbers peaks in Fig. 4. The maximum number of photons which should be possible to distinguish using this device is $k_m = \Delta T_C / \Delta T_e^{h\nu}$ where $\Delta T_e^{h\nu} = h\nu/C_e \approx 6$ mK at $T = 100$ mK is the initial increase of the electron temperature caused by an absorbed photon. Based on this argument, $k_m$ = 3-4 should be possible. Indeed, a weak 3-photon peak can be seen in Figs. 3 and 4. A peak with $k = 4$ should not be seen since the ratio $h_4/h_0$ with $\mu = 0.47$ is ~ 0.002. An increase of $\mu$ quickly shifts the entire histogram towards larger pulse amplitudes and to the decrease of the peak contrast.

The $I(T_e)$ curve in Fig. 5 predicts a maximum current sensitivity $dI/dT_e$ = 1.8 nA/mK at 100 mK that would result in a photon sensitivity of 11 nA/photon. In the experiment, the distance between photon number peaks in Fig. 4 is just 4 nA/photon. If the effective energy scale in Fig. 4 were adjusted by a factor of $\varepsilon$ = 4/11 = 0.36, the noise which is added after the photon energy loss, would be just $\Delta E_{FWHM}$ = 39 meV, in good agreement with the value predicted by model, $\Delta E_{FWHM}$ = 38 meV.

Other factors increasing $\Delta E_{FWHM}$ apparently played a lesser role. The table value of $\gamma$ fits well to the present work's data as well as to the measured values of $\tau_{e\text{-}ph}$ found both directly and through the ratio of $C_e/G_{e\text{-}ph}$ [13,16]. Also the noise originating from the fluctuation of the number of high-energy phonons escaping to the substrate [26] did not seem to be significant here since the width of the photon-number peaks is described well by the detector intrinsic noise (Fig. 4).

The positive shift of the zero-photon peak of histograms in Figs. 4 and 5 for large values of $\mu$ is an indication that there are some low energy events creating an increase of the measured current (for $k = 0$, no photons are absorbed in the HEB but all photons still land in other areas). Those events could be a result of various photon energy "downconversion" processes not originating in the Ti sensor (e.g., emission of phonons with broad energy spectrum after a photon absorption in the Nb leads or/and Si). These processes did not add noise as the photon number peak widths remain fairly constant even for large $\mu$ values (Fig. 3).

In conclusion, we have demonstrated a calorimetric detection of single mid-IR photons with $\lambda = 8$ μm using a Ti superconducting hot-electron microbolometer. The achieved $\Delta E_{FWHM}$ is consistent with the theory when a reasonable photon energy conversion efficiency $\varepsilon \approx 36\%$ is taken into account. The results are very promising with potential applications in astronomical spectral imaging of faint sources [12,13], free-space quantum communication [27], and single-molecule spectroscopy [28] Smaller devices are technologically feasible that should result in an improved energy resolution and also allow for efficient optical coupling using lithographic microantennas [29,30]. Submicron size devices should allow for single-photon detection in the THz range, where possible issues with generation of phonons in the substrate can be avoided due to the small absorption of the THz radiation in Si.

We thank A.G. Kozorezov and A.V. Sergeev for discussion and comments and J. Kawamura for providing the superconducting solenoid. The research was carried out at the Jet Propulsion Laboratory, California Institute of Technology under a contract with the National Aeronautical and Space Administration. The work at Yale University was supported in part by NSF-DMR-0907082, NSF-CHE-0616875, and Yale University. The research at Rutgers University was supported in part by the National Aeronautics and Space Administration (NASA) grant NNG04GD55G, the Rutgers Academic Excellence Fund, and the NSF grant ECS-0608842.

1   C. K. Stahle, D. McCammon, and K. D. Irwin, Phys. Today **52**, 32 (1999).

2   K. D. Irwin and G. C. Hilton, in *Cryogenic Particle Detection* (Springer-Verlag, Berlin, 2005), Vol. 99, pp. 63.

3   B. Cabrera and R. W. Romani, in *Cryogenic Particle Detection* (Springer-Verlag, Berlin, 2005), Vol. 99, pp. 417.

4   B. Cabrera, R.M. Clarke, P. Colling, A.J. Miller, S. Nam, and R.W. Romani, Applied Physics Letters **73**, 735 (1998).

5   A.E. Lita, A.J. Miller, and S.W. Nam, Optics Express **16**, 3032 (2008).

6   M.D. Petroff, M.G. Stapelbroek, and W.A. Kleinhans, Applied Physics Letters **51**, 406 (1987).

7   S. Komiyama, IEEE Journal of Selected Topics in Quantum Electronics **17**, 54 (2011).

8   I. Milostnaya, A. Korneev, M. Tarkhov, A. Divochiy, O. Minaeva, V. Seleznev, N. Kaurova, B. Voronov, O. Okunev, G. Chulkova, K. Smirnov, and G. Gol'tsman, Journal of Low Temperature Physics **151**, 591 (2008).

9   A. J. Miller, S. W. Nam, J. M. Martinis, and A. V. Sergienko, Applied Physics Letters **83**, 791 (2003).

10  D. Rosenberg, A. E. Lita, A. J. Miller, and S. W. Nam, Phys. Rev. A **71** (2005).

11  R. H. Hadfield, Nat. Photonics **3**, 696 (2009).

12  B. S. Karasik and A. V. Sergeev, IEEE Transactions on Applied Superconductivity **15**, 618 (2005).

13  J. Wei, D. Olaya, B. S. Karasik, S. V. Pereverzev, A. V. Sergeev, and M. E. Gershenson, Nature Nanotechnology **3**, 496 (2008).

14  D. F. Santavicca, B. Reulet, B.S. Karasik, S.V. Pereverzev, D. Olaya, M.E. Gershenson, L. Frunzio, and D. E. Prober,  Applied Physics Letters **96**, 083505 (2010).

15  B.S. Karasik, S.V. Pereverzev, D. Olaya, M.E. Gershenson, R. Cantor, J.H. Kawamura, P.K. Day, B. Bumble, H.G. LeDuc, S.P. Monacos, D.G. Harding, D. Santavicca, F. Carter, and D.E. Prober,  Proceedings of SPIE **7741**, 774119 (2010).

16  B.S. Karasik, A.V. Sergeev, and D. E. Prober,  IEEE Transactions on Terahertz Science and Techonlogy **1**, 97 (2011).

17  M. E. Gershenson, D. Gong, T. Sato, B. S. Karasik, and A. V. Sergeev,  Applied Physics Letters **79**, 2049 (2001).

18  A. Soibel, K. Mansour, G. Spiers, and S. Forouhar,  MRS Proceedings **883**, FF2.5 (2005). [doi:10.1557/PROC-883-FF2.5]

19  D.J. Benford, M.C. Gaidis, and J.W. Kooi,  Applied Optics **42**, 5118 (2003).

20  C. Gmachl, A. Tredicucci, F. Capasso, A. L. Hutchinson, D. L. Sivco, J. N. Baillargeon, and A. Y. Cho,  Applied Physcs Letters **72**, 3130 (1998).

21  K. D. Irwin,  Applied Physics Letters **66**, 1998 (1995).

22  B. S. Karasik and A. I. Elantiev,  Applied Physics Letters **68**, 853 (1996).

23  S. H. Moseley, J. C. Mather, and D. McCammon,  Journal of Applied Physics **56**, 1258 (1984).

24  S.W. Nam, B. Cabrera, P. Colling, R. M. Clarke, E. Figueroa-Feliciano, A.J. Miller, and R.W. Romani,  IEEE Transactions on Applied Superconductivity **9**, 4209 (1999).

25  A.E. Lita, A.J. Miller, and S. Nam,  Journal of Low Temperature Physics **151**, 125 (2008).

[26] A.G. Kozorezov, J.K. Wigmore, D. Martin, P. Verhoeve, and A. Peacock, Applied Physics Letters **89**, 223510 (2006).

[27] G. Temporao, H. Zbinden, S. Tanzilli, N. Gisin, T. Aellen, M. Giovanni, J. Faist, and J.P. von der Wied, Quantum Information and Computation **8**, 1 (2008).

[28] Y. L. A. Rezus, S. G. Walt, R. Lettow, A. Renn, G. Zumofen, S. Gotzinger, and V. Sandoghdar, Physical Review Letters **108**, 093601 (2012).

[29] J. Alda, J. M. Rico-García, J. M. López-Alonso, and G. D. Boreman, Nanotechnology **16**, S230 (2005).

[30] P. Biagioni, J.-S. Huang, and B. Hecht, Reports on Progress in Physics **75**, 024402 (2012).

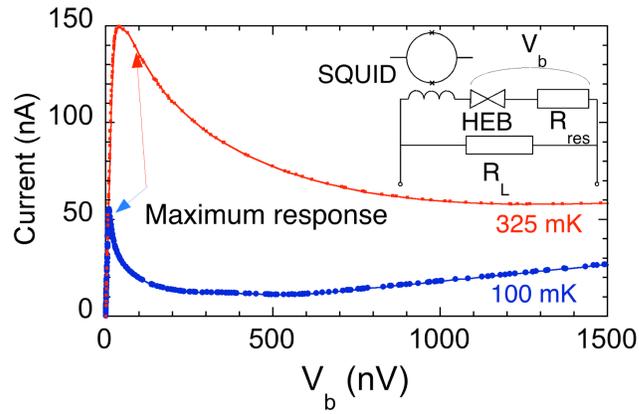

Fig. 1. Current-voltage characteristics (IVC) of the HEB device at 325 mK (zero magnetic field, $B = 0$) and 100 mK ($B > 0$). A linear slope at low voltage bias is due to the residual resistance (normal metal connections between the HEB and the bias circuit), $R_{res} = 0.14$ Ω. The IVC peak current (150 nA @ 325 mK and 58 nA @ 100 mK) roughly corresponds to the critical current, $I_C$. Single photon detection at 100 mK was observed only in the bias range corresponding to the resistive state (i.e., $V_b > 30$ nV). An inset shows the circuit diagram.

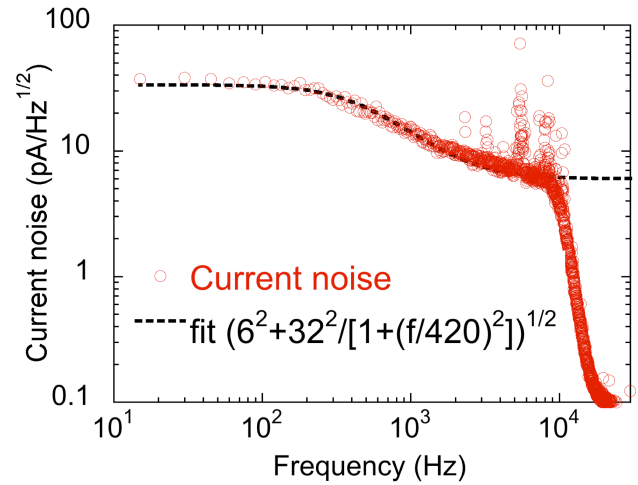

Fig. 2. Noise spectrum of the HEB device at 100 mK. The dotted line shows the frequency dependence, with two plateaus. This is typical for a TES. The sharp cutoff at ~ 10 kHz is due to an external low-pass filter. Below ~ 400 Hz, the noise shown by the fitted line is a sum of the TEF and the Johnson and SQUID noise. Above 400 Hz, the TEF noise rolls off and only the Johnson and the SQUID noise remain.

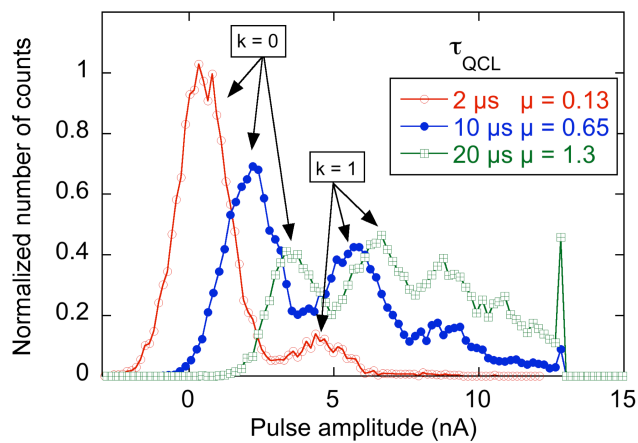

Fig. 3. Amplitude histogram for the HEB device at 50 mK with different QCL pulse durations. $\tau_{QCL}$ sets the average number of absorbed photon per pulse, $\mu$. For these three cases $\mu = 0.13$, 0.65, and 1.3. The labeled photon number peaks, $k = 0, 1$, are shown for all three values of $\mu$.

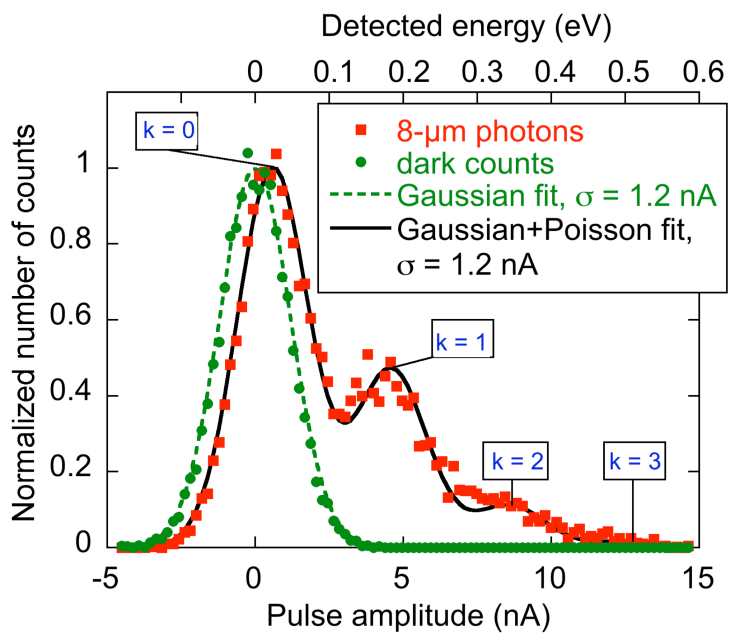

Fig. 4. Photon count histogram for the HEB device at 100 mK. Both the dark count and photon count statistics fit well with a Gaussian function with an rms deviation $\sigma = 1.2$ nA. The solid line

is the modeling of the count statistics using a combination of the Poisson and Gaussian distributions. An average number of absorbed photons per pulse in the Poisson distribution $\mu$ = 0.47. The labels show photon number peaks $k$ = 0, 1, 2, and 3.

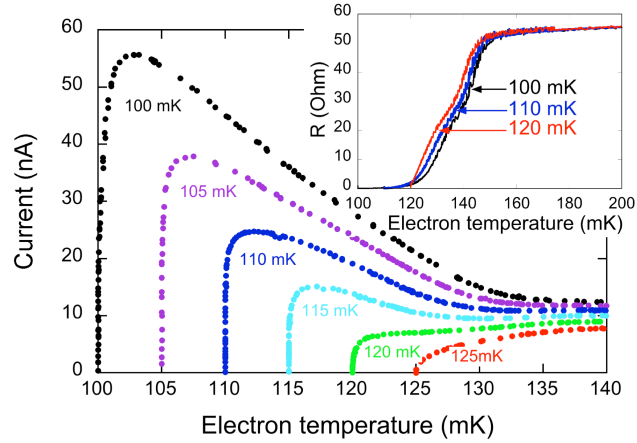

Fig. 5. Dependence of the current on the electron temperature derived from experimental IVCs and a heat-balance equation (Eq. 2). Each curve corresponds to a different bath temperature $T$ (labeled by the curve). A nearly linear variation of the current vs. $T_e$ for $T$ = 100 mK explains the total number of the observed photon number peaks and their equidistant positions (see Fig. 4). An inset shows $R(T_e)$ curves recovered from the same data set (the labels indicate the bath temperature for each curve). They almost coincide thus indicating the validity of the thermal model assuming that the bolometer is a lumped element and all its characteristics can be described by an empirical $R(T_e)$ dependence and Eq. 2.